\documentclass[preprint,12pt]{elsarticle}

\usepackage{amssymb}
\usepackage{amsthm}
\usepackage{amsmath,amssymb,amsfonts}%
\usepackage{multirow}%
\usepackage{tabularx}
\usepackage{caption}
\usepackage{subcaption}
\usepackage{float}
\usepackage{multicol}
\usepackage{multirow}
\usepackage{booktabs}
\usepackage{caption}
\usepackage[utf8]{inputenc}
\usepackage{xcolor}
\usepackage{soul}
\usepackage{listings}%
\usepackage{threeparttable}
\usepackage{textcomp}%
\usepackage{xcolor}%
\usepackage{mathrsfs}%
\usepackage[polish,english]{babel} 
\usepackage[colorlinks,citecolor=red,urlcolor=blue,bookmarks=false,hypertexnames=true]{hyperref}
\usepackage[export]{adjustbox} 

\journal{Acta Materialia}

\begin{document}
\begin{frontmatter}



\title{Rapid modeling of segregation-driven metal-oxide adhesion in high-entropy alloys using macroscopic atom model}


\author[a]{Dennis Boakye}
\author[a]{Chuang Deng \corref{b}}
\cortext[b]{Corresponding author}
\ead{chuang.deng@umanitoba.ca}

\affiliation[a]{organization={Mechanical Engineering, University of Manitoba},
            addressline={66 Chancellors Cir}, 
            city={Winnipeg},
            postcode={R3T 2N2}, 
            state={Manitoba},
            country={Canada}}

\begin{abstract}
Accurate prediction of metal–oxide adhesion in high-entropy alloys (HEAs) is challenging because interfacial segregation, atomic environments, and macroscopic thermodynamic quantities are strongly correlated. Relying solely on first-principles approaches is too expensive for exploring composition, solute concentration, and co-segregation effects. To address this, we extend the macroscopic atom model (MAM) for multicomponent alloys using composition-consistent surface fractions and an interfacial pair-probability formalism that captures deviations from random contact statistics. Applied to CoCrFeNi (AlCoCrFeNi) HEA in contact with $\mathrm{Cr_2O_3}$ ($\mathrm{Al_2O_3}$), the model predicts segregation energies and work of separation as continuous functions of composition, reproducing the correct segregation hierarchy of Hf, Y, Zr, and S. The stronger segregation tendency at $\mathrm{Al_2O_3}$ interfaces, and the non-linear dependence of surface energy and adhesion on solute content and co-segregation is also captured. The results are benchmarked with DFT calculations, which shows consistent trends, particularly the strengthening of adhesion by Hf and Zr through strong metal–oxygen bonding and the weakening effect of S. These results demonstrate that the extended MAM provides a physically interpretable, computationally efficient, and quantitatively predictive framework for screening segregation-controlled adhesion beyond the limits of DFT.
\end{abstract}




\end{frontmatter}

\section{Introduction}\label{intro}
The durability of an oxide scale is determined not only by its growth rate, but also by its ability to remain attached to the underlying alloy \cite{fontana1967corrosion,birks2006introduction}. Failure of this attachment, through spallation or interfacial weakening by trace impurities, often governs the lifetime of the protective scale in service. Predicting impurity segregation and its effect on interfacial adhesion requires a framework that links local chemical driving forces to macroscopic thermodynamic quantities. Such a framework must remain practical across wide compositional spaces and dilute solute concentrations. Experiments on superalloys \cite{babic2020reactive,evans2011oxidation,smialek1987effect} and high-entropy alloys (HEAs) \cite{holcomb2015oxidation,HUANG2024174597,liu2023comparative} show that trace elements, such as Sulfur (S), can strongly reduce adhesion, while Yttrium (Y) and Hafnium (Hf) can mitigate this effect and restore adhesion even at very low concentrations. However, experimental observations are inherently system-specific and rarely give a transferable quantitative measure of why a given impurity segregates or how segregation alters adhesion energy across different alloy compositions. First-principles calculations, such as density functional theory (DFT), provide an atomistic view of bonding and segregation energetics but become expensive for larger systems and multiple competing solutes \cite{jiang2008first,boakye2024reactive,jiang2016efficient}. This creates a gap between atomistic understanding and predictive capability, especially when segregation tendencies and adhesion trends are composition-dependent.  

To bridge this gap, semi-empirical models, such as the macroscopic atom model (MAM) \cite{niessen1989macroscopic,de1988cohesion}, translate atomic-scale interface physics into meaningful quantitative thermodynamic estimates. The model treats an alloy as built from atomic cells whose properties are determined by the pure elements in the metallic state, and relates energy effects mainly to the interfaces between dissimilar neighboring atomic cells \cite{niessen1989macroscopic}. The energetic consequences of alloying can be described as the atomic-scale analog of bringing two macroscopic metal blocks into contact \cite{niessen1989macroscopic,rohrer2001structure}. This creates discontinuity at the interface, and the energy cost (or gain) of eliminating them can be used to determine related thermodynamic quantities.

According to the model's description of metal-metal contact, two elemental parameters are paramount: the electron density $n_{ws}$ and an adjusted work-function-like quantity, $\phi^*$ \cite{niessen1989macroscopic}. $n_{ws}$ is defined as the electron density at the boundary of the Wigner-Seitz-type atomic cell in the pure metal. When different cells are brought into contact, a mismatch in $n_{ws}$ results in discontinuities in electron density that must be smoothed, generating a positive contribution to the interfacial energy \cite{niessen1989macroscopic,rohrer2001structure}. $\phi^*$ captures the driving force for net charge transfer across a metal-metal interface. This charge transfer results in an interfacial dipole layer that negatively contributes to the formation enthalpy, reflecting the energy gain associated with the transfer of charge until the contact potential is compensated \cite{niessen1989macroscopic}.

As such, two competing physical effects, density mismatch (penalty) and charge transfer (stabilization), arise from $n_{ws}$ and $\phi^*$, respectively, and combine to produce the characteristic MAM form for alloying enthalpies. According to the formulation \cite{niessen1989macroscopic,de1988cohesion}, the sign and magnitude of formation tendencies can be expressed using a compact competition between a term scaling with $(\Delta\phi^*)^2$ and a term scaling with $(\Delta n_{ws})^2$, which explains why some binary combinations strongly form compounds while others phase-separate. Using the same interfacial physics, the surface, interfacial, and adhesion energies can be calculated, since these quantities are determined by the creation or elimination of the contact area between atomic cells and their neighbors. For example, the model explicitly treats a metal-vacuum interface as an abrupt discontinuity in the electron density, thereby enabling the estimation of surface energies of pure metals \cite{niessen1989macroscopic,de1988cohesion}.

The MAM framework has been applied to predict the heats of mixing \cite{yuan2015interface,de1988cohesion}, surface energies \cite{ossi1988surface,yan2013size,de1988cohesion}, formation enthalpies \cite{raju2016estimation}, adsorption energies \cite{niessen1989macroscopic}, and segregation enthalpies in dilute alloys \cite{de1988cohesion}. In particular, the framework connects segregation propensity to the tendency of low-surface-energy species to enrich at surfaces and to the tendency of species with unfavorable solution enthalpies to reduce unfavorable unlike contacts by segregating. This exact combination helps treat competing solutes, such as S, and elements, such as Y and Hf, where interfacial bonding and chemical interactions with bulk elements drive segregation even at trace levels \cite{holcomb2008calculation,pint1995reactive,fritscher2023reactive,liu2023comparative}.  

The MAM provides a consistent thermodynamic path from elemental parameters to interfacial energy changes during segregation, which is difficult to reproduce experimentally or by DFT. Although experimental studies can identify which species segregate and correlate these trends with scale adhesion or spallation behavior \cite{pint1996experimental,moon1989role,holcomb2008calculation,liu2023comparative}, they do not directly produce the interfacial energy landscape required to explain trends across different alloy compositions. In contrast, DFT can provide quantitative segregation energies and work of separation for specific atomic configurations \cite{jiang2008first,boakye2024reactive,lan2014effects}, but its application becomes prohibitive for systematic sampling of solute concentration, multiple competing solutes, and chemical disorder, particularly when the objective is to explore alloy families rather than isolated interface models \cite{friedel1954electronic}. In this paper, MAM offers a complementary technique by retaining physically interpretable driving forces while remaining computationally efficient enough to explore broad composition spaces and support alloy-design reasoning \cite{vitos1994full,niessen1989macroscopic,de1988cohesion}.

Despite its success in elemental and dilute alloy systems, extending MAM to multi-principal element alloys has not yet been established. Classical implementations of the model typically assume a dominant solvent matrix with dilute solute additions \cite{niessen1989macroscopic,de1988cohesion}, an assumption that breaks down in HEAs, where each constituent acts simultaneously as a solvent and a solute and contributes comparably to surface and interfacial energetics \cite{cantor2004microstructural,yeh2004nanostructured,miracle2017critical}. In such systems, segregation frequently involves competition and co-segregation among multiple elements rather than a single solute–solvent pair \cite{middleburgh2014segregation,ye2016elemental,ferrari2020surface}, complicating both the definition of host properties and the estimation of effective contact statistics. These challenges must be addressed before MAM can be used as a predictive tool for segregation and adhesion in chemically complex alloys, rather than as a qualitative analogy.

In this study, we extend the framework to predict segregation and adhesion in multicomponent alloys, with a focus on HEA compositions relevant to oxidation-resistant systems \cite{lu2020hf,holcomb2015oxidation,liu2023comparative}. We apply the same interfacial physics while introducing a compositionally consistent treatment of local environments appropriate to multicomponent hosts. The model serves as the primary predictive tool for determining segregation hierarchies and their implications for adhesion. Finally, DFT calculations are used to validate key energetic trends and provide an atomistic interpretation of the predicted driving forces. The study aims to provide a scalable and physically grounded approach to predicting segregation-mediated trends in HEAs beyond the scope of exhaustive DFT calculations.

\section{Theory and calculations}
\subsection{Interfacial adhesion}
The work of separation $\mathrm{W_{sep}}$ is defined as the reversible energy required to separate a unit area of the oxide/metal interface from its free surfaces \cite{finnis1996theory}. Figure  \ref{fig:adhesion} shows that the cleavage energy required to separate two surfaces in contact is higher than $\mathrm{W_{sep}}$.  The work of adhesion, $F_{ad}$, is the change in reversible free-energy to obtain free surfaces from an interface. It is apparent that $F_{ad}$ is lower than $\mathrm{W_{sep}}$ due to the presence of surface impurities. The higher $\mathrm{W_{sep}}$, the greater the energy required for cleavage. The ideal $\mathrm{W_{sep}}$  can be expressed by Dupr\'e equation \cite{finnis1996theory}:
\begin{figure}[t]
    \centering
    \includegraphics[width=0.8\linewidth]{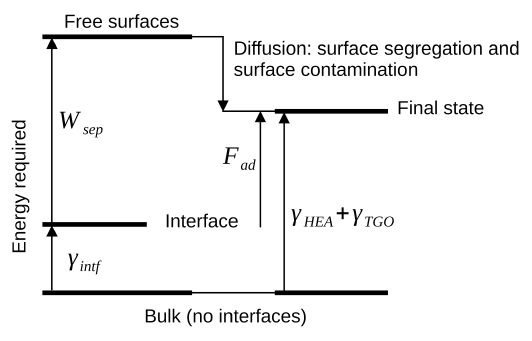}
    \caption{A diagram of the energies involved in interface cleavage, which excludes plastic processes \cite{finnis1996theory}.}
    \label{fig:adhesion}
\end{figure}

\begin{equation}
  \mathrm{W_{sep}} \;=\; \gamma_{\mathrm{HEA}} \;+\; \gamma_{\mathrm{TGO}} \;-\; \gamma_{\mathrm{int}},
  \label{eq:wad}
\end{equation}
where $\gamma_{\mathrm{HEA}}$ and $\gamma_{\mathrm{TGO}}$ are the excess surface energies of the HEA and the oxide film, respectively, and $\gamma_{\mathrm{int}}$ is the excess interfacial energy of the joined system in units of $\mathrm{Jm^{-2}}$. A larger $\mathrm{W_{sep}}$ implies stronger bonding and therefore higher resistance to interfacial decohesion.

In HEAs, the surface energy $\gamma$ arises from the collective contribution of all alloying species. $\gamma$ can be computed using \cite{bennett2005modeling}:
\begin{equation}
    \gamma_0 = \frac{1}{f_{\text{vac}}c_0}\sum^n_i\frac{C^S_i\Delta H^{surf}_i}{V^{2/3}_i} \label{eq:surf_en}  
\end{equation}
where $n$ is the number of elements in the alloy or TGO, $C_i^S$ is the surface fraction of the component $i$, $V_i$ is the corrected molar volume, and the constant $c_0$ is given by $4.5\times 10^8\,\mathrm{mol}^{-1/3}$ \cite{de1988cohesion}. $f_{vac}=0.31$ is the degree to which the atoms on the surface are exposed to vacuum, and $\Delta H^{\mathrm{surf}}_i$ is the surface enthalpy of $ith$ element, which is tabulated in Table S2 of the supplemental material and in ref. \cite{de1988cohesion}. For O and S, $\Delta H^{\mathrm{surf}}$ is not tabulated and can be estimated using Eq. \eqref{eq:surf_en} and the surface energy of $\mathrm{Al_2O_3}$ and $\mathrm{FeS_2}$, respectively.  

The surface fraction $C_i^S$ of element $i$ is initially estimated to be proportional to its atomic fraction $c_i$ weighed by its molar volume $V_i$ \cite{de1988cohesion}:
\begin{equation}
  C_i^S \;=\;
  \frac{c_i V_i^{2/3}}{\sum_{k=1}^{n} c_k V_k^{2/3}}. 
  \label{eq:surf_frac}
\end{equation}

In multicomponent systems, charge transfer can change effective atomic volumes. To capture this, we sequentially pick a matrix element $X$ and treat all others as minor elements. The volume of the matrix element in the alloy $(V_X)_{\text{alloy}}$ is further corrected using a valence-dependent factor \cite{de1988cohesion}.
\begin{equation}
  \left(V_X^{2/3}\right)_{\mathrm{alloy}}
  = \left(V_X^{2/3}\right)_{\mathrm{pure}}
  \left[1 + a f_X\left(\phi_X^* - \bar{\phi}^*\right)\right],
  \label{eq:vol_corr}
\end{equation}
where $a$ is the valence factor (0.07 for trivalent/noble metals such as Al, 0.10 for divalent metals such as Ni, and 0.04 for others, including O and S) \cite{de1988cohesion,neuhausen2003extension}. The parameter $f_X$ is a contact factor that describes the geometric surroundings of the atom $X$, $\phi^*_X$ is the Miedema electronegativity of $X$, and $\bar{\phi}^*$ is the average Miedema electronegativity of all minor elements given as
\begin{equation}
  \bar{\phi}^* \;=\; \frac{\sum\limits_{i\neq X} c_i^{S_0}\,\phi_i}
                         {\sum\limits_{i\neq X} c_i^{S_0}}.
\end{equation}
Here, $c_i^{S_0}$ is the initial surface fraction calculated using Eq. \eqref{eq:surf_frac}. 

In contrast to the classical implementation of $f_X$ \cite{de1988cohesion,niessen1989macroscopic}, in HEAs, interface energetics depend on the statistics of nearest-neighbor contacts \cite{zhao2021role}, which can deviate sharply from the mean-field approach due to local chemical fluctuations or short-range order. Therefore, we redefine $f_X$ in terms of normalized interfacial pair probabilities as
\begin{equation}
    f_X = \sum\limits_{j\neq X}\frac{P_{Xj}^{int}}{C^S_X} \label{eqn:fx}
\end{equation}
where $P_{Xj}^{int}$ is the probability of forming an interfacial contact between species $X$ and $j$. $P_{Xj}^{int}$ can be written as
\begin{equation}
    P_{Xj}^{int} = C^S_X\frac{C^S_j\left(1 - \alpha^{int}_{Xj}\right)}{\sum_kC^S_k\left(1 - \alpha^{int}_{Xk}\right)} \label{eqn:pint}
\end{equation}
where $\alpha^{int}_{ij}$ accounts for deviations from random contact statistics.  If $\alpha^{int}_{ij} > 0$, the $ith$ and $jth$ elements avoid each other and vice versa. For random-mixing limit, $\alpha^{int}_{ij}=0$,  Eq. \eqref{eqn:fx} reduces to
\begin{equation}
    f_X = \sum\limits_{j\neq X}C^S_j = 1 - C^S_X
\end{equation}
which is consistent with the original implementation that only the fraction of unlike neighbors contributes to the contact factor \cite{niessen1989macroscopic,de1988cohesion}. The final surface fraction can be iteratively evaluated using the results of Eq. \eqref{eq:vol_corr} to recompute $C^S_i$ from Eq. \eqref{eq:surf_frac}, which is then used to calculate the surface energy in Eq. \eqref{eq:surf_en}. 

$\gamma_{\text{int}}$ in Eq. \eqref{eq:wad} can be defined as the sum of a chemical term and a mismatch term, where the contribution of the mismatch term can be approximated by analogy with a high-angle grain boundary  given by \cite{de1988cohesion}:
\begin{equation}
  \gamma^{\mathrm{mis}} \;=\;
  \tfrac{1}{3}\left(\frac{\gamma_{\mathrm{HEA}} + \gamma_{\mathrm{TGO}}}{2}\right).
  \label{eq:mis}
\end{equation}

The chemical term, $\gamma^{chem}$, comprises enthalpic and entropic contributions. At $\mathrm{0\,K}$, the entropic contribution can be neglected, and $\gamma_{\text{int}}$ is purely determined by the interaction enthalpy between the atoms on the surface. The interaction enthalpy $\Delta H_{ij}^\circ$ is defined as the change in enthalpy when a mole of the solute $i$ is dissolved in an infinite solution of solvent $j$ and has the unit of $\mathrm{kJ/mol}$ \cite{de1988cohesion,bennett2005modeling}.
\begin{equation}
    \Delta H_{ij}^\circ = \frac{V^{2/3}_i}{\left(n^{-1/3}_{ws}\right)_{av}}\left[-P\left(\Delta\phi^*\right)^2 + Q\left(\Delta n^{1/3}_{ws}\right)^2 \right]
    \label{eq:ent}
\end{equation}
where $P$ and $Q$ are experimentally determined constants and the value of $\mathrm{Q/P=9.4\,V^2/\text{density units}^{2/3}}$. $V_i$ is the corrected molar volume calculated using Eq. \eqref{eq:vol_corr} in $\mathrm{cm^3/mol}$, $\Delta\phi^*$ is the difference in the chemical potential derived from the work function of the pure elements in V, and $\Delta n_{ws}$ is the difference in the electron density at the boundary of the Wigner-Seitz cell in density units \cite{de1988cohesion}. From Eq. \eqref{eq:ent}, $\Delta H_{ii}^\circ = \Delta H_{jj}^\circ = 0$ for similar components across the interface. The values of $\phi$, $n_{ws}$ and $V$ are tabulated in Table S1 of the supplemental material and in ref. \cite{de1988cohesion} for transition metals and ref. \cite{neuhausen2003extension} for S and O.

Using Eqs. \eqref{eq:ent} and \eqref{eq:surf_frac}, the chemical term can be written as:
\begin{equation}
    \gamma^{chem}_{HEA\rightarrow TGO} = \sum\limits_{i=1}^m\sum\limits_{j=1}^nP^{\text{int}}_{ij}\frac{\Delta H_{A_iB_j}^\circ}{c_0V^{2/3}_{A_i}} \label{eq:chem}
\end{equation}
where we have considered that the interface of HEA and TGO has elements $A_i = 1, ..., m$ and $B_j = 1, ... n$, respectively. $P^{int}_{ij}$ accounts for the probability of pair interaction between $i-j$ at the interface.

From Eq. \eqref{eq:chem}, $\gamma^{int}$ is calculated from the perspective of the HEA. A similar approach can be adopted to shift the perspective to TGO, giving $\mathrm{\gamma^{int}_{TGO\rightarrow HEA}}$. The average of these two is taken so that $\gamma^{int}$ reflects the molar volumes of both surfaces:
\begin{equation}
    \gamma^{int} = \tfrac{1}{2} \left(\gamma^{chem}_{\mathrm{HEA\rightarrow TGO}} + \gamma^{chem}_{\mathrm{TGO\rightarrow HEA}}\right) +   \tfrac{1}{6}\left(\gamma_{\mathrm{HEA}} + \gamma_{\mathrm{TGO}}\right).
\end{equation}

\subsection{Interfacial segregation}
The interfacial segregation of solute elements is driven by differences in chemical bonding between the solute, the alloy matrix, and the oxide. Within MAM, this driving force is described using dilute-limit enthalpies of solution \cite{de1988cohesion}. 

For a solute atom A segregating to an HEA-TGO interface, only a fraction of its nearest neighbors lie across the interface; for close-packed metallic interfaces, this fraction is taken as $\alpha=1/3$ \cite{de1988cohesion}. Using Eqns. \eqref{eqn:pint} and \eqref{eq:ent} the heat of segregation $\Delta H^{int}_{seg}(A)$ can be written as
\begin{equation}
\begin{split}
        \Delta H^{int}_{seg}(A) = &\alpha\left[\sum_jC_j^{TGO}\Delta H_{A\in j}^\circ -  \sum_i\sum_jP^{int}_{ij}\Delta H_{i\in j}^\circ - \right. \\ &\left. \sum_iC^{HEA}_i\Delta H_{A\in i}^\circ \right] + \Gamma_{HEA} 
        \label{eq:seg}
\end{split}
\end{equation}
where $C_i^{HEA}$ and $C_j^{TGO}$ are the surface fractions of the $ith$ and $jth$ elements in the HEA and oxide, respectively. The last term on the right vanishes for an epitaxial interface. $\Gamma_{HEA}$ measures the mismatch of the lattices and is given by \cite{de1988cohesion}:
\begin{equation}
    \Gamma = 0.05c_0V^{2/3}_A\left(\gamma_A - \sum_iC^S_i\gamma^{HEA}_i\right)
\end{equation}
where $\gamma_A$ is the surface energy of the solute, $\gamma_i$ is the surface energy of the $ith$ element in the HEA, and all other symbols have their usual meaning as defined before. 

\subsection{Geometry and DFT calculations}
\begin{figure}
    \centering
    \includegraphics[width=0.8\linewidth]{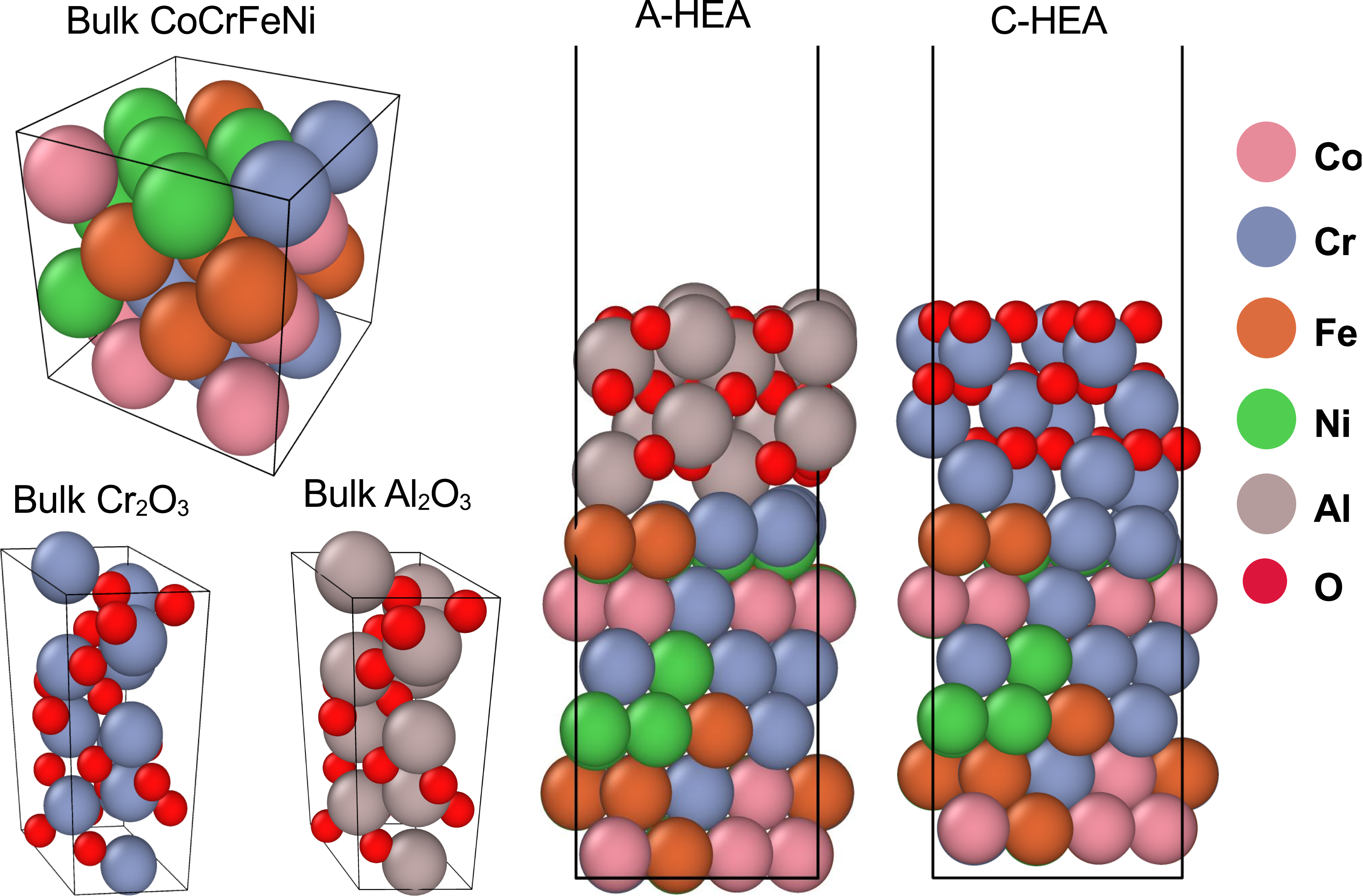}
    \caption{Perspective view of relaxed bulk CoCrFeNi HEA, bulk chromia ($\mathrm{Cr_2O_3}$), bulk alumina ($\mathrm{Al_2O_3}$), and side views of chromia-HEA (C-HEA) and alumina-HEA (A-HEA) interfaces.}
    \label{fig:geometry}
\end{figure}

The bulk structure of the HEA was generated using the Special Quasirandom Structure (SQS) approach \cite{zunger1990special} implemented in the Alloy Theoretic Automated Toolkit (ATAT) \cite{van2002alloy}, which provides a statistically representative description of chemical disorder in multicomponent alloys. A $2\times2\times2$ supercell containing 32 atoms was used for bulk calculations. Crystalline structures ($R\bar{3}c$) of alumina and chromia were obtained from the Materials Project database \cite{jain2013commentary} and used to construct the TGO components of the interface models. The alloy-TGO slab geometries were generated with a vacuum spacing of 15 \r{A} to prevent spurious interactions between periodic images. All interface supercells contained 156 atoms, with the bottom 3 layers fixed to mimic the bulk and the top layers allowed to relax. As shown in Figure \ref{fig:geometry}, all structural visualization and post-processing were performed using the OVITO software package \cite{stukowski2009visualization}.

The DFT calculations were then performed using the Vienna ab initio Simulation Package (VASP) \cite{hafner2008ab} within the projector augmented-wave (PAW) formalism \cite{kresse1999ultrasoft}. Exchange–correlation effects were described using the Perdew–Burke–Ernzerhof (PBE) \cite{perdew1996generalized} form of the generalized gradient approximation (GGA). A plane-wave kinetic energy cutoff of 500 eV was used throughout, after ensuring convergence. Structural relaxations were performed until the total energy convergence criterion of $10^{-5}$ eV was satisfied, and the residual Hellmann-Feynman forces in all atoms were reduced below 30 meV/\r{A}. Standard valence electron configurations as recommended by VASP for each element were used. Electronic occupancies were treated using the Methfessel–Paxton smearing scheme \cite{methfessel1989high} with a smearing width of 0.2 eV. Brillouin zone sampling was carried out using Monkhorst–Pack \cite{monkhorst1976special} k-point meshes of $4\times4\times4$ for bulk structure and $3\times3\times1$ for slab and interface models, consistent with the reduced periodicity along the surface normal. Spin polarization was enabled to account for the magnetic contributions of the transition-metal elements in the alloys.

\section{Results and discussions}
\subsection{Interfacial segregation mechanism} \label{sect:segregation}
\subsubsection{Thermodynamic driving forces for segregation}
The segregation of trace impurities to metal-oxide interfaces constitutes the first significant step in interfacial cohesion and scale adhesion. Since only species that occupy interfacial sites can influence interfacial bonding, it is imperative to establish both the tendency for segregation and its physical origin.  

\begin{figure}[t]
    \centering
    \includegraphics[width=1\linewidth]{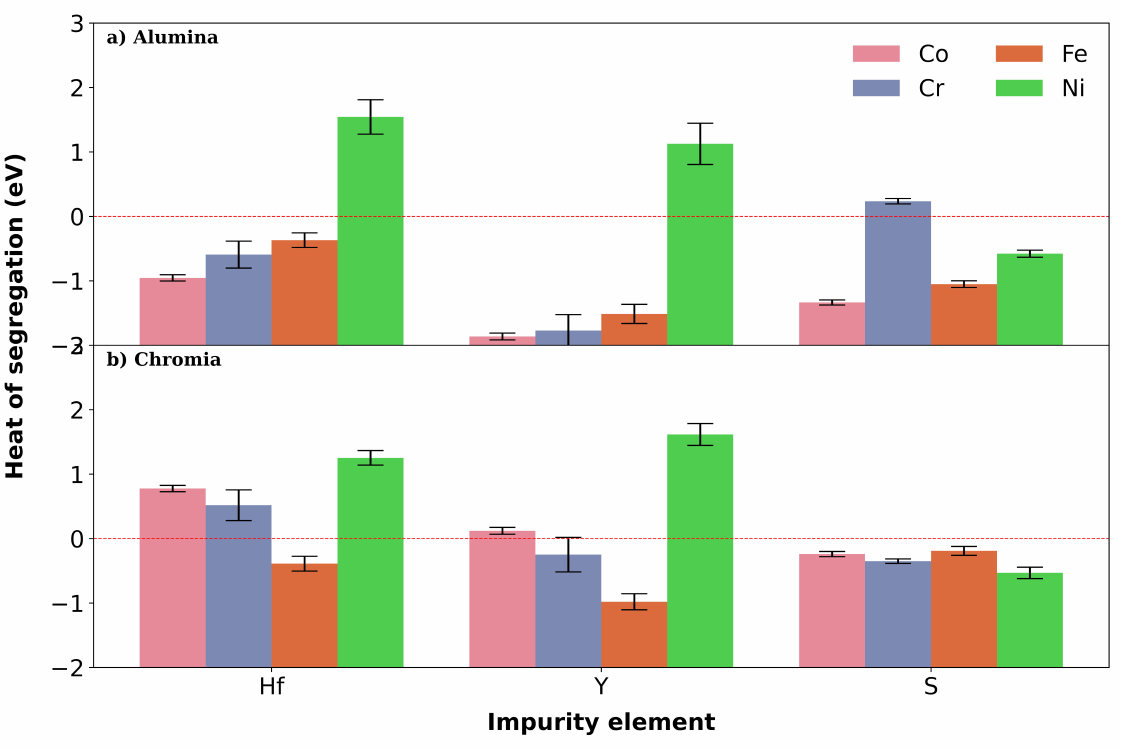}
    \caption{Segregation of trace elements (Hf, Y, and S) in (a) alumina, and (b) chromia interfaces. The segregation tendency (negative) of Y and S is higher at the alumina interface than at the chromia interface. Notice how host Fe sites readily make segregation possible for REs.}
    \label{fig:segregation_dft}
\end{figure}

Figure \ref{fig:segregation_dft} summarizes the DFT-calculated segregation energies of selected trace impurities at the $\mathrm{CoCrFeNi/Al_2O_3}$ and $\mathrm{CoCrFeNi/Cr_2O_3}$ interfaces. Negative segregation energies indicate a thermodynamic preference for the solutes to occupy interfacial sites relative to the bulk alloy. The segregation tendency of Y is higher compared to that of both Hf and S. In this context, Y will outcompete S for interfacial sites, while S will outcompete Hf. For example, segregation to Fe sites is favorable for all solutes in both interfaces, with Y having the highest tendency followed by S. This theoretical observation agrees with the works of Liu et al \cite{liu2023comparative}. The SIMS study revealed S segregation in the Hf-doped NiCoCrAl, but no segregation in the Y-doped alloy. Co-doping of Y with Hf suppressed S segregation, improving oxidation resistance. Moreover, the tendency for S to segregate to Ni outcompetes that of Hf and Y, which do not segregate. Therefore, interfacial Ni vacancies can be a sink for S accumulation. This could potentially explain why Ni-superalloys are susceptible to S degradation during high-temperature oxidation. 

Another significant observation in Figure \ref{fig:segregation_dft} is that the tendency of trace elements to segregate in the alumina interface is higher compared to the chromia interface. This enhanced segregation tendency reflects the higher ionicity and stronger metal-oxygen bonding associated with alumina. Nevertheless, such strong segregation tendencies have been experimentally observed to facilitate oxygen diffusion, leading to the inward growth of oxides and, subsequently, internal oxidation \cite{hou1995effect,boakye2025effect,jedlinski1993comments,jedlinski1993general}. In the chromia interface, the segregation tendency of RE is lower, while still higher than that of S, especially in the case of Y doping. This synergistic effect of moderate segregation of REs and lower segregation of S makes $\mathrm{Cr_2O_3}$ suitable for maximizing the effect of REs in highly oxidizing environments. The availability of Ni- and Co-interfacial host sites can facilitate S segregation at the chromia interface, whereas those of Hf and Y, which are more aggressive at Fe sites, cannot. A study by Lu et al. \cite{lu2020hf} in $\mathrm{AlCoCrFeNi}$ HEA showed S segregation in the Fe-free sample even with appreciable concentrations of Hf and Y.  Hence, in the absence of Fe, the tendency to segregate is limited and becomes ineffective in outcompeting S.

\subsubsection{Chemical origin of RE segregation}
\begin{figure}[t]
    \centering
    \includegraphics[width=1\linewidth]{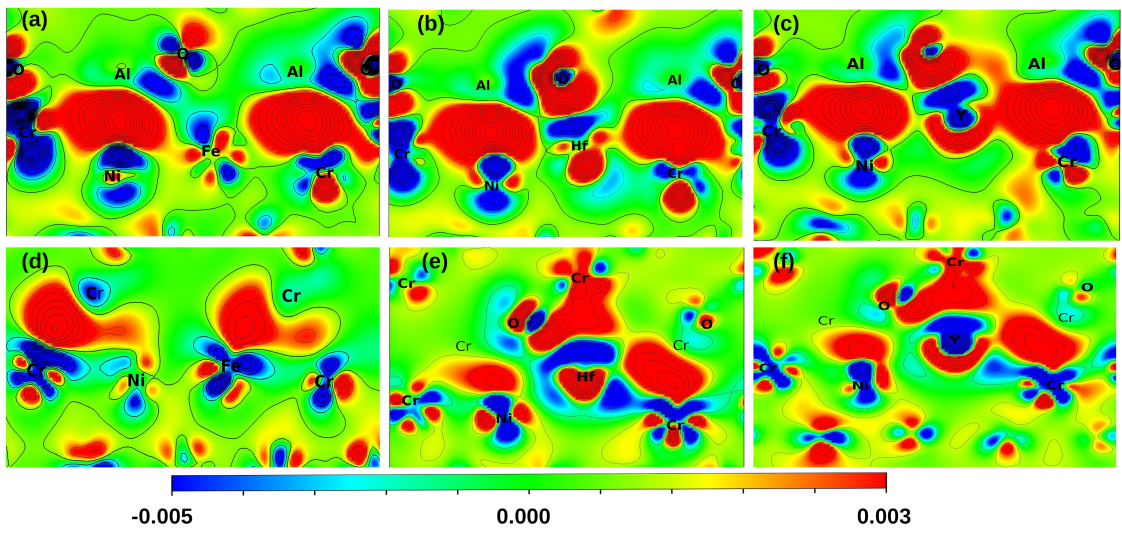}
    \caption{2D view of the charge-density difference for (a) and (d) no doping, (b) and (d) Hf-doping, and (c) and (d) Y-doping, of alumina- and chromia interfaces, respectively.}
    \label{fig:charge}
\end{figure}

Although the segregation energies in Figure \ref{fig:segregation_dft} establish that REs segregate, we provide an explanation of why they do so using the electronic structure at the interface. In Figure \ref{fig:charge}, the 2D difference in charge-density for the Hf and Y interfaces reveals a pronounced charge accumulation along the RE-O bonds, accompanied by charge depletion around neighboring metallic atoms. This redistribution is significantly stronger than that observed for matrix-element-oxygen bonds, indicating that segregation is primarily driven by chemically strong RE-O interactions rather than by elastic relaxation alone.

This interpretation, driven by bonds, is quantitatively explained by the values of $\Delta H_{ij}^\circ$ between the elements. For example, the dilute-limit enthalpies for the Hf-O (-1113 kJ/mol) and Y-O (-953 kJ/mol) interactions are significantly more exothermic than those for the Ni-O (-241 kJ/mol), Co-O (-302 kJ/mol), Cr-O (-578 kJ/mol) or Al-O pairs (-838 kJ/mol) \cite{de1988cohesion}. Thus, replacing a matrix atom at the interface with an RE leads to a net reduction in the chemical interaction energy. From Eq. \eqref{eq:seg}, this stabilization arises predominantly from the solute-oxide interaction term, which outweighs both the average matrix-oxide interaction and the solute's interaction with the bulk alloy environment. Hence, REs experience a strong thermodynamic driving force to migrate from bulk to oxygen-rich interfacial sites. In contrast to REs, S segregation originates from a different mechanism. Although S also lowers interfacial free energy, it does so primarily by weakening the metallic bonding rather than by forming strong directions with O, as was observed in our previous work \cite{boakye2024reactive}. This process explains the well-known detrimental influence of S on oxide-scale adhesion and provides a thermodynamic basis for its antagonistic action with REs.

\subsubsection{MAM-based generalization to HEAs}
\begin{figure}[tb]
    \centering
    \includegraphics[width=1\linewidth]{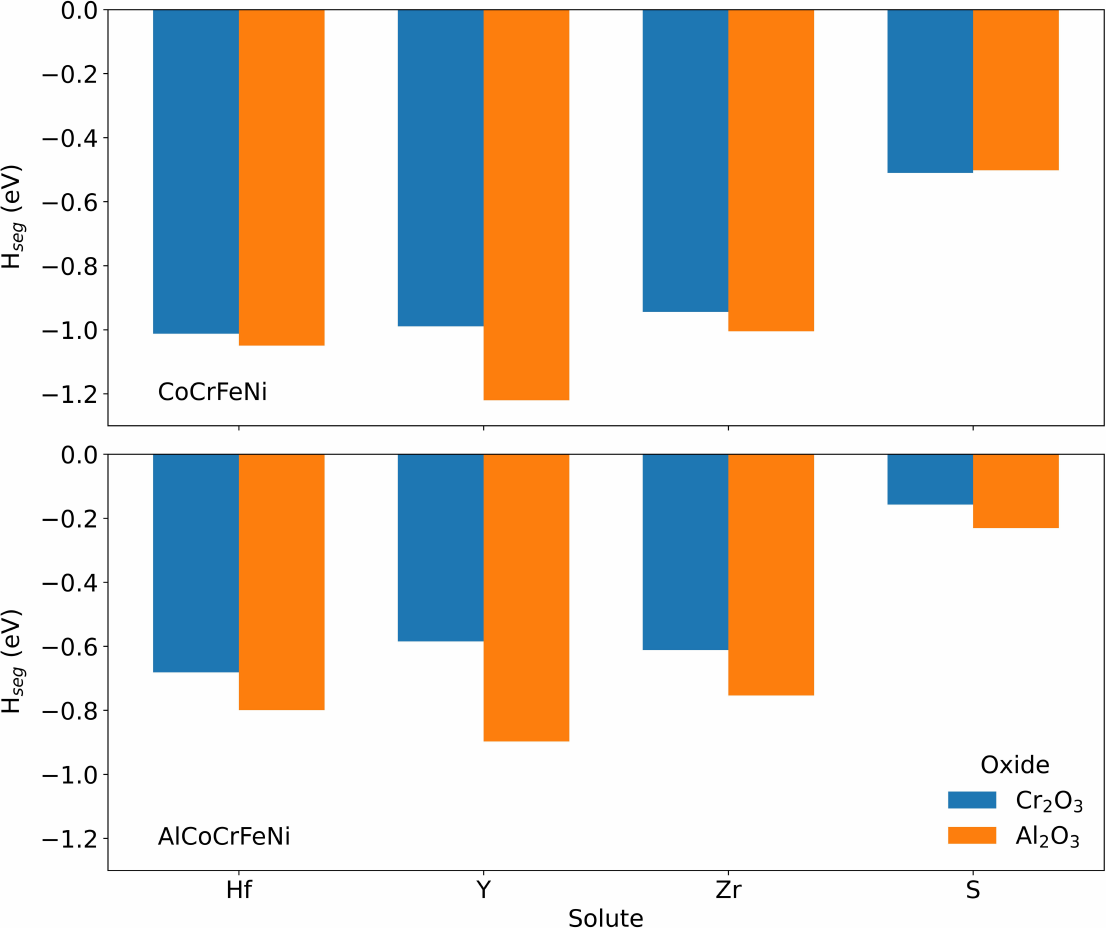}
    \caption{Segregation of trace elements (Hf, Y, Zr, and S) in CoCrFeNi and AlCoCrFeNi HEAs. All REs have a higher propensity to occupy interfacial sites and outcompete S.}
    \label{fig:seg_mam}
\end{figure}

To extend our results beyond atomistic first-principles, we employed the MAM to evaluate segregation tendencies across multiple alloy chemistries and impurity species. In Figure \ref{fig:seg_mam}, the segregation energies predicted by MAM reproduce the hierarchy derived from DFT for CoCrFeNi, capturing the strong preference of Y and Hf for interfacial sites relative to S. Significantly, MAM enables the systematic exploration of other impurities such as Zr and more complex HEAs, including AlCoCrFeNi, where DFT data are missing.

One other thing to note is that although there are differences in absolute magnitudes between the DFT and MAM calculations, the consistency in relative trends confirms that segregation is predominantly governed by chemical interaction energies rather than by system-specific structural details. Figure \ref{fig:seg_mam} shows that the tendency for RE segregation is higher in the alumina interface compared to the chromia interface, which is consistent with the DFT results. Moreover, both Y and Hf have a higher segregation tendency than S and will therefore outcompete S at both interfaces. This also agrees with the DFT results and our previous findings \cite{boakye2024reactive}. The extension of the model to Zr, a typical RE that is usually used to improve oxidation resistance \cite{smialek1987effect,li2014influence}, indicates that not only Hf and Y are capable of pinning S, but also other REs. In comparison, Zr's tendency to segregate is lower than that of Y and Hf. Hence, the MAM provides a general framework for accessing the segregation behavior of trace elements in HEAs.

\subsection{Surface thermodynamics by segregated solutes (MAM)}
The surface energy plays a central role in determining both the stability of free surfaces and the energetic cost of forming oxide-metal interfaces. The surface energy of the clean CoCrFeNi surface was independently calculated using DFT to be approximately 2.5 $\mathrm{Jm^{-2}}$, which is in agreement with the predicted 2.45 $\mathrm{Jm^{-2}}$. This increases confidence in the model's ability to capture physically meaningful trends in surface energy. 

\begin{figure}
    \centering
    \includegraphics[width=1\linewidth]{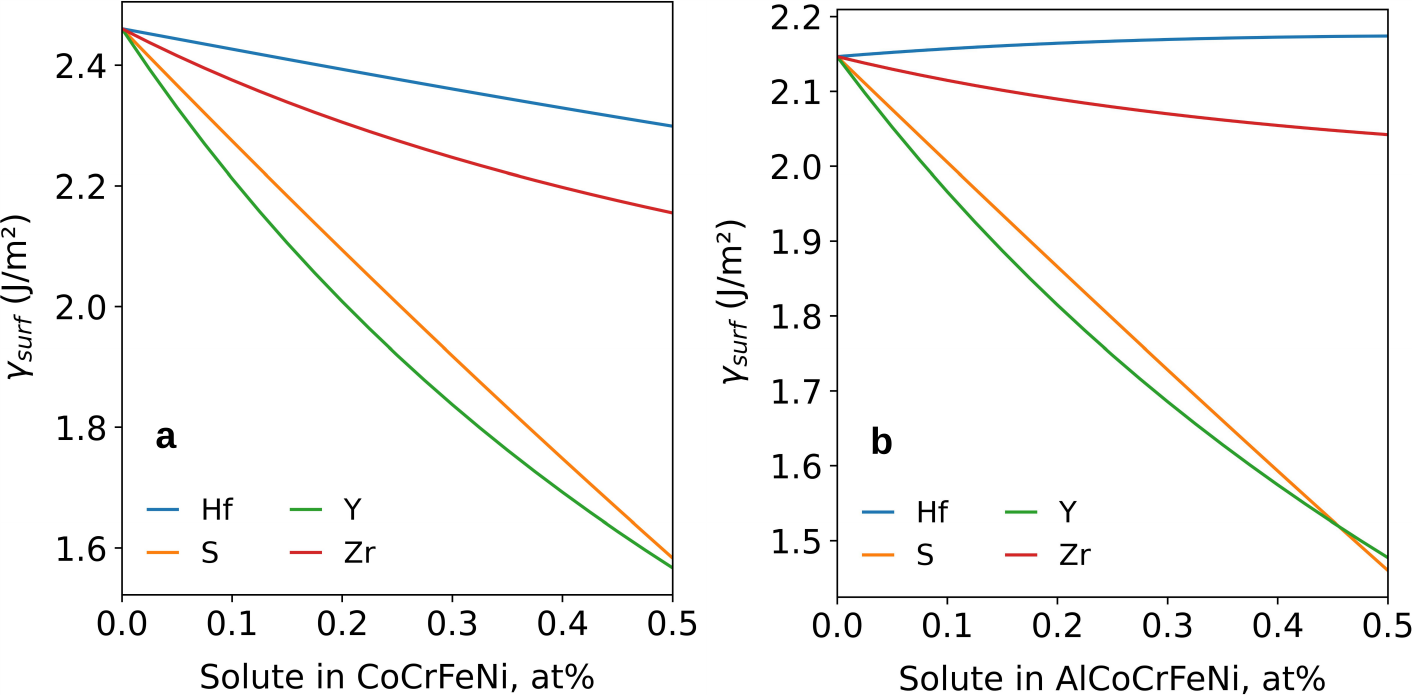}
    \caption{Effect of trace impurities on the surface energy of (a) CoCrFeNi and (b) AlCoCrFeNi HEAs. Notably, both Y and S lower the surface energy, while the reverse is the case for Hf and Zr.}
    \label{fig:surf_co_al}
\end{figure}

Figure \ref{fig:surf_co_al} shows the variation of the CoCrFeNi and AlCoCrFeNi HEAs with the concentration of solutes when trace elements are introduced. Except for S, the surface energy responds nonlinearly to solute concentration, reflecting the competing influences of chemical interactions and geometric effects at the surface.  Reactive elements generally increase surface energy at low to intermediate concentrations. This increase arises from the strong chemical affinity of REs for oxygen, which stabilizes interfacial bonding but simultaneously raises the energetic cost of maintaining a free metallic surface. In contrast, unlike Hf, both Zr and Y reduce surface energy, with Y producing the greatest effect. This is mainly due to the low surface enthalpy of Y compared to Hf and Zr. Moreover, S tends to reduce surface energy over a broad concentration range because its surface enthalpy is significantly lower than that of REs. This behavior is consistent with the tendency of sulfur to weaken metallic bonding and reduce surface cohesion, an effect long associated with its detrimental influence on the adhesion of the oxide scale \cite{pint1995reactive,liu2023comparative}.

Furthermore, the magnitude of these effects strongly depends on the chemistry of the base alloy. Compared with CoCrFeNi, the Al-containing HEA exhibits a more pronounced response to RE additions, reflecting the altered electronic environment and bonding characteristics introduced by Al. Moreover, the surface enthalpy of Al (76 kJ/mol) is significantly lower compared to Cr (121 kJ/mol), Co (127 kJ/mol), Fe (129 kJ/mol) and Ni (121 kJ/mol) \cite{de1988cohesion}. These differences explain the importance of explicitly considering alloy chemistry when assessing the thermodynamic consequences of segregation in multicomponent systems.

\begin{figure}[t]
    \centering
    \includegraphics[width=1\linewidth]{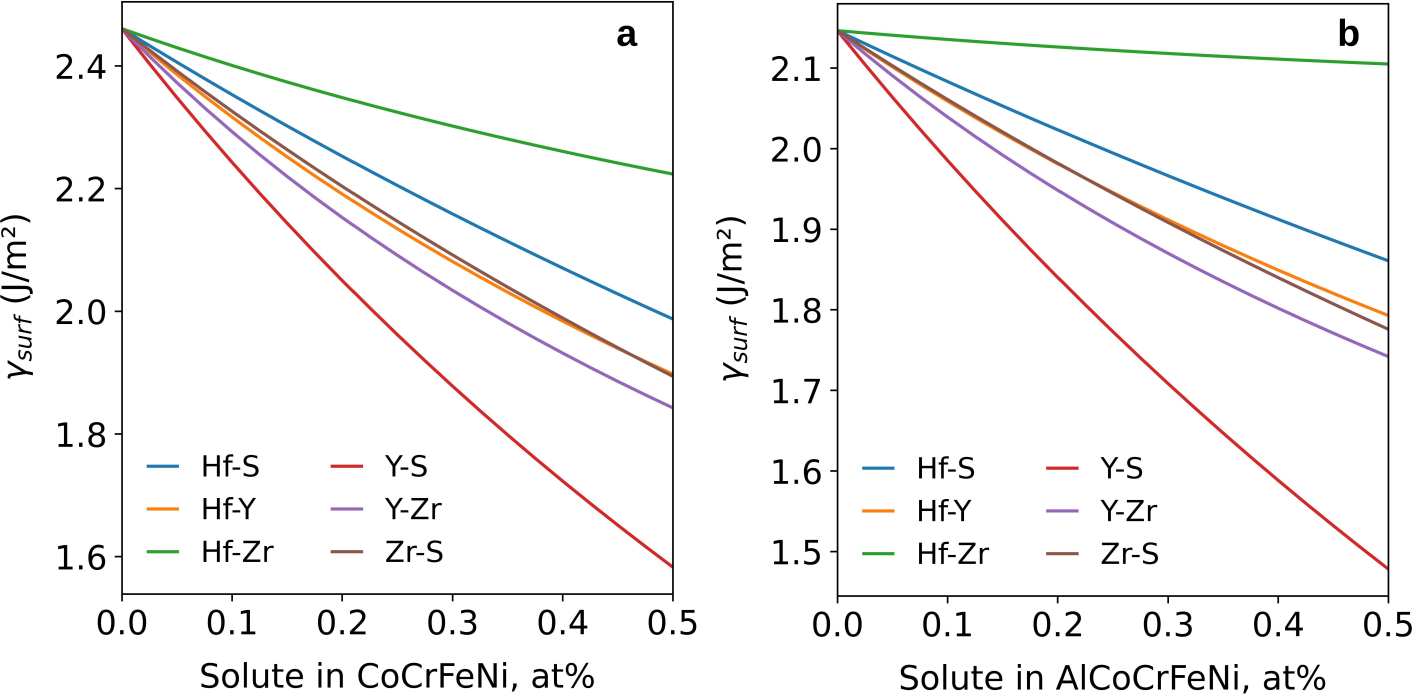}
    \caption{Effect of trace impurities on the surface energy of (a) CoCrFeNi and (b) AlCoCrFeNi HEAs. Notably, both Y and S lower the surface energy, while the reverse is the case for Hf and Zr.}
    \label{fig:surf_pair}
\end{figure}
Figure \ref{fig:surf_pair} illustrates the effect of co-segregation on the surface energies of CoCrFeNi and AlCoCrFeNi HEAs. It is apparent that an RE-S combination has strongly non-additive behavior. The surface energy of the Y-S combination is the lowest and decreases with a combined concentration. This is obvious since the surface enthalpies of both Y and S are smaller than those of Hf and Zr. The least reduction in surface energy occurred for the co-segregation of Hf-Zr in the AlCoCrFeNi HEAs. In contrast, Hf-Zr only slightly decreases the surface energy with concentration in the AlCoCrFeNi HEA. In several cases, the increase in RE segregation partially offsets the reduction induced by S, leading to moderate surface energies that cannot be achieved through a single-solute addition alone. In reality, solute concentrations such as those discussed in this work rarely reach 1\% composition; extending to 50\% helps illustrate their effect at higher concentrations. Moreover, the observed nonlinearity reflects direct chemical interactions among solutes as well as changes in local coordination and surface composition. This behavior demonstrates that the assumption of a dilute, single-solute environment is insufficient for complex engineering alloys, which typically contain multiple coexisting trace elements.

\subsection{Effect of segregated species on interfacial adhesion}
\subsubsection{Atomistic benchmarks for adhesion (DFT)}
As mentioned earlier, the mechanical integrity of TGOs is ultimately determined by the adhesion of the oxide to the underlying alloy substrate. Extensive experimental evidence has demonstrated that scale spallation during thermal cycling is rarely controlled by bulk mechanical properties alone, but rather by chemical modifications at the oxide–metal interface arising from the segregation of trace elements \cite{hou1995effect,pint1995reactive,liu2023comparative}. In this context, $W_{\text{sep}}$ provides a physically meaningful measure of interfacial cohesion that links the atomic-scale bonding to the macroscopic failure phenomena observed in oxidation experiments \cite{evans2011oxidation,jiang2008first}.
\begin{figure}[tb]
    \centering
    \includegraphics[width=1\linewidth]{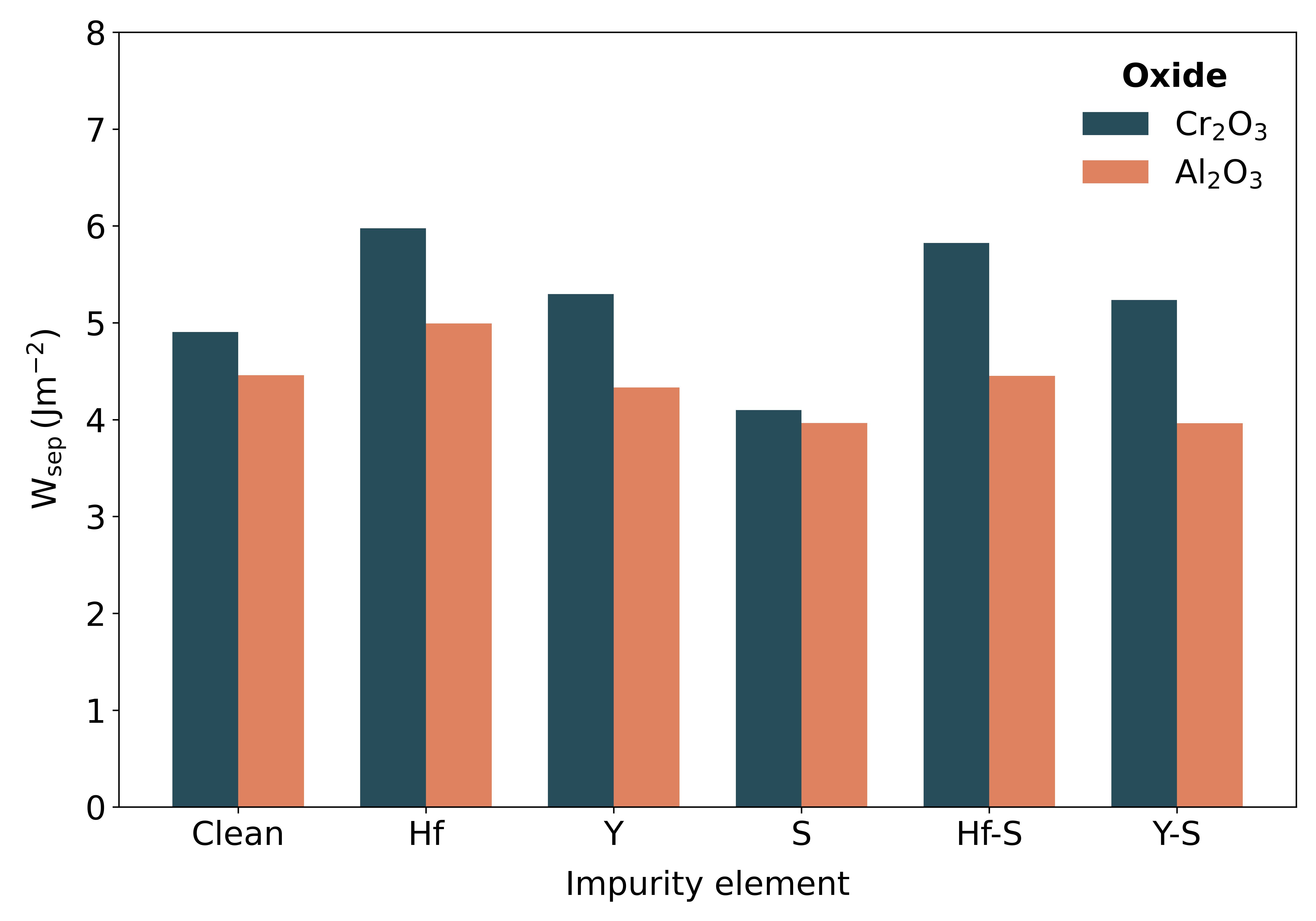}
    \caption{Effect of trace elements (Hf, Y, and S) on the interfacial adhesion of
    alumina ($Al_2O_3$)-, and chromia ($Cr_2O_3$)-forming CoCrFeNi HEAs from DFT. Notice the detrimental effect of S and increased adhesion in REs.}
    \label{fig:wsep_dft}
\end{figure}
    
DFT calculations for the $\mathrm{CoCrFeNi/Al_2O_3}$ and $\mathrm{CoCrFeNi/Al_2O_3}$ interfaces show that REs and S exert opposing influences on scale adhesion. As shown in Figure \ref{fig:wsep_dft} for $\mathrm{Cr_2O_3}$, the segregation of Hf or Y at the interface leads to a substantial increase in $Wsep$ from 4.9 $\mathrm{Jm^{-2}}$ in the clean interface to 6.0 $\mathrm{Jm^{-2}}$ with Hf and 5.2 $\mathrm{Jm^{-2}}$ with Y, while sulfur causes a reduction to 4.05 $\mathrm{Jm^{-2}}$. Moreover, the effect of S is slightly higher in the $Al_2O_3$ interface compared to the $\mathrm{Cr_2O_3}$ interface. These trends closely reproduce long-established experimental observations that trace sulfur dramatically degrades oxide scale adhesion \cite{hou1995effect,pint1996experimental}, while additions of REs such as Y, Hf, and Zr markedly improve adherence in alloys forming both chromia and alumina \cite{smialek1987effect,liu2023comparative,li2014influence}. In $\mathrm{Al_2O_3}$, the adhesion of the clean interface is slightly reduced, showing the ineffectiveness of Y compared to Hf. 

The atomistic origin of these adhesion trends is directly related to the segregation behavior discussed in Section \ref{sect:segregation}. Reactive elements segregate strongly at oxide–metal interfaces and form strong bonds with interfacial oxygen, as evidenced by the pronounced charge density distribution in Figure \ref{fig:charge} and in ref. \cite{boakye2024reactive}. These strong RE–O bonds increase the energetic cost of interfacial decohesion, increasing $W_{sep}$. In contrast, S segregation weakens interfacial cohesion by disrupting metallic bonds and reducing the effective strength of the metal-oxygen bond \cite{jiang2008first,boakye2024reactive}. Experimental surface-sensitive techniques, including AES and SIMS, have shown that sulfur can reach very high interfacial concentrations even when present at bulk levels below tens of parts per million \cite{liu2023comparative,holcomb2015oxidation,lu2020hf}, providing a clear explanation for its disproportionately severe impact on scale adhesion and embrittlement.

The co-segregation of S with the REs provides further understanding of the mechanisms controlling adhesion. In Figure \ref{fig:wsep_dft}, the DFT results indicate that the presence of Hf or Y partially restores $W_{sep}$ at both interfaces containing S and Hf, the latter exhibiting the strongest mitigating effect. This behavior reflects competition for interfacial sites and the ability of REs to re-establish strong bonding pathways across the interface. Moreover, the effect of RE is more pronounced in the $\mathrm{Cr_2O_3}$ interface compared to the $\mathrm{Al_2O_3}$ interface, confirming the effectiveness of REs in chromia-forming alloys \cite{hou1995effect}.

These findings agree with experimental observations that REs do not always completely eliminate S from oxide–metal interfaces but, nonetheless, significantly improve scale adhesion \cite{pint1995reactive,hou1995effect}. Experimentally, S has been reported to remain detectable at interfaces even in RE-doped alloys, yet spallation resistance is markedly enhanced \cite{holcomb2015oxidation}. The present results provide a thermodynamic explanation for this observed behavior. Rather than acting solely as sulfur getters, REs shift the interfacial chemistry balance toward cohesion, counteracting the deleterious effects of S.

\subsubsection{Effect of single- and co-segregation on adhesion (MAM)}
\begin{figure}[h!]
    \centering
    \includegraphics[width=1\linewidth]{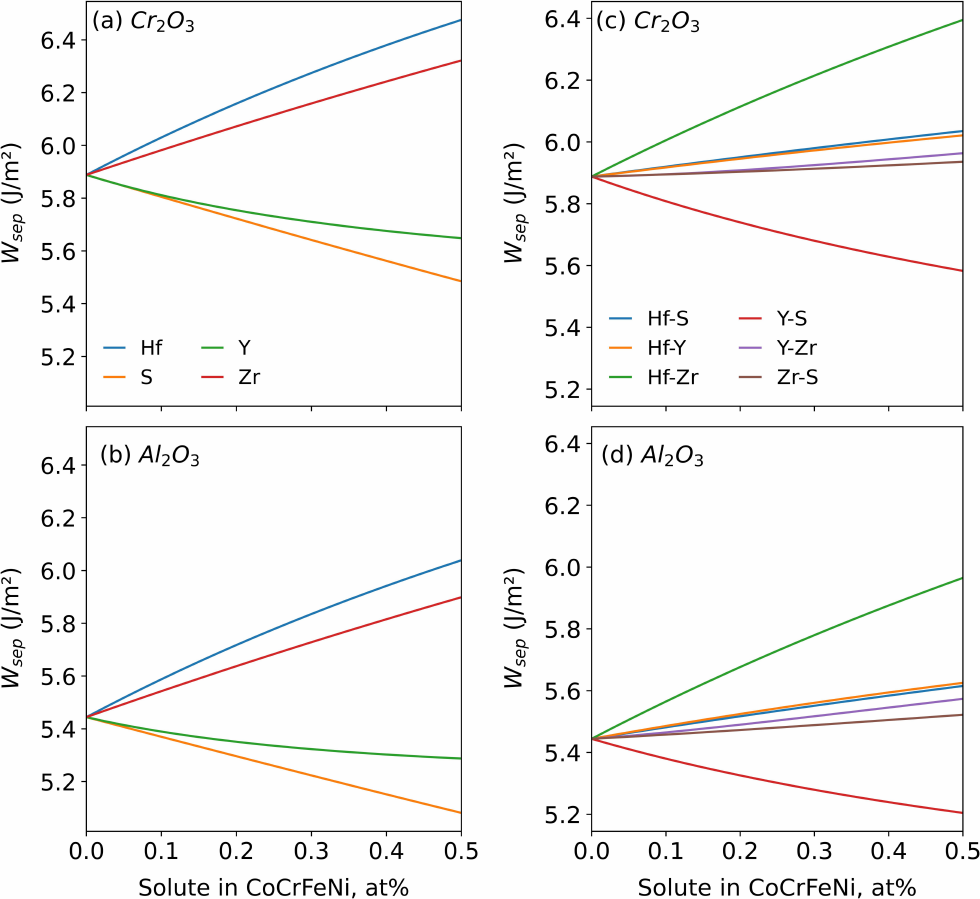}
    \caption{Effect of trace impurities on the surface energy of (a) CoCrFeNi and (b) AlCoCrFeNi HEAs. Notably, both Y and S lower the surface energy, while the reverse is the case for Hf and Zr.}
    \label{fig:wsep_co_al}
\end{figure}

The MAM extends these atomistic insights by enabling systematic exploration of adhesion as a function of solute concentration and co-segregation. The concentration-dependent $W_{sep}$ trends shown in Figure \ref{fig:wsep_co_al} exhibit strong non-linear behavior, particularly at low solute levels. Even small additions of REs lead to disproportionately large increases in adhesion, whereas sulfur causes a rapid decline, consistent with our DFT calculations. In single solute segregation, Hf and Zr produce the highest $W_{sep}$ in all range of solute concentrations, while Y slightly reduces adhesion. In the case of co-segregation, Hf-Zr exhibits the largest $W_{sep}$, followed by Hf-Y in the AlCoCrFeNi HEA and Hf-S in the CoCrFeNi HEA. Although all RE co-segregate positively with S, Y-S produces the worst adhesion. Notwithstanding, in all concentration ranges of solutes, the $W_{sep}$ of Y-S is higher than the single solute effect of S. 

This nonlinearity provides a thermodynamic explanation for a long-standing industrial observation: REs are effective at extremely low bulk concentrations because strong segregation amplifies their influence at the interface. Nevertheless, such is not the case, since the segregation of REs to grain boundaries inhibits the diffusion of cations but facilitates anion diffusion \cite{jedlinski1993comments,jedlinski1993general,boakye2025effect}. Similar arguments have been advanced in the oxidation studies of alloys that form alumina, where ppm-level additions of Y or Hf produce dramatic improvements in resistance to cyclic oxidation and scale adhesion \cite{hou2008segregation,HUANG2024174597}.

\begin{figure}[tb]
    \centering
    \includegraphics[width=1\linewidth]{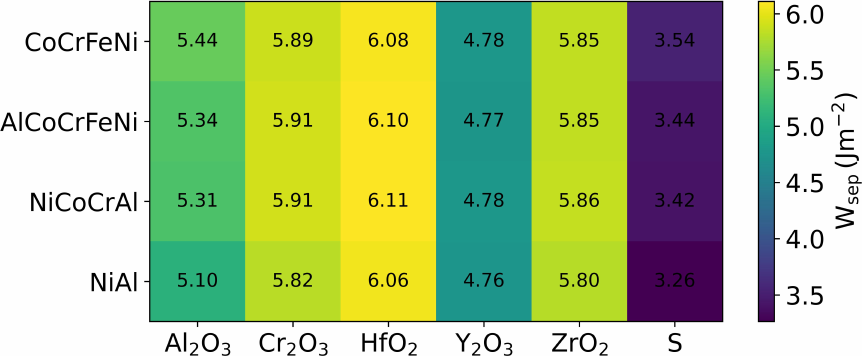}
    \caption{Estimation of $W_{sep}$ when REs are added as oxide dispersoids as compared to TGOs and S. Notice the positive effect of $HfO_2$ and $ZrO_2$ on $W_{sep}$ and the negative effect of S.}
    \label{fig:wsep_disp}
\end{figure}

As shown in Figure \ref{fig:wsep_disp}, the composition of an alloy also modulates the adhesion response. Compared to CoCrFeNi, Al-containing alloys exhibit a lower $W_{sep}$, reflecting changes in electronic structure and bonding introduced by Al. This observation is consistent with experimental findings that RE effects differ between chromia-forming and alumina-forming systems and that chromia scales are often more sensitive to beneficial segregants. The AlCoCrFeNi and CoCrFeNi alumina interface exhibits slightly lower $W_{sep}$ of 5.34 and 5.44 $\mathrm{Jm^{-2}}$, respectively, while that of chromia is 5.91 and 5.89 $\mathrm{Jm^{-2}}$, respectively. The NiCoCrAl chromia interface produces results similar to those of AlCoCrFeNi HEA. The lowest adhesion is observed for the NiAl alumina interface with a $W_{sep}$ of 5.10 $\mathrm{Jm^{-2}}$. The NiAl alumina interface is of practical importance due to its application in turbine engines \cite{grabke1999oxidation,jayne1993sulfur,jayne2024sulphur,ohtsu2007oxidation}. Nevertheless, alumina formation is preferred due to reactive evaporation of chromia at elevated temperatures \cite{holcomb2008calculation}.  

Reactive elements can be added as alloy additions or oxide dispersoids. In the latter case, these dispersoids form interfaces with the matrix. Because these oxides are stable, they do not break up to form alloy additions. However, improved scale adhesion has been reported in dispersoid-added alloys \cite{pint1995reactive, nagai1981effect}. The broader trends in adhesion across oxide dispersoids are summarized in Figure \ref{fig:wsep_disp}. $HfO_2$ exhibits the highest adhesion, 6.11 $\mathrm{Jm^{-2}}$ followed by $ZrO_2$. In fact, $Y_2O_3$ reduces the adhesion of the clean interface for both alumina and chromia in all alloy compositions. $W_{sep}$ for S for all alloy compositions is drastically reduced compared to clean interfaces. It is well established that S is highly detrimental in NiAl alloys \cite{hou2008segregation}, which requires the use of thermal barrier coatings \cite{evans2011oxidation,vv2019role,chen2017mechanistic}. These results align with industrial strategies that employ REs and oxide dispersoids to stabilize protective scales and extend the lifetimes of components in aggressive high-temperature environments, including turbine components, heating elements, and protective coatings.

\subsubsection{Effect of chemical ordering on adhesion}
\begin{figure}[tb]
    \centering
    \includegraphics[width=1\linewidth]{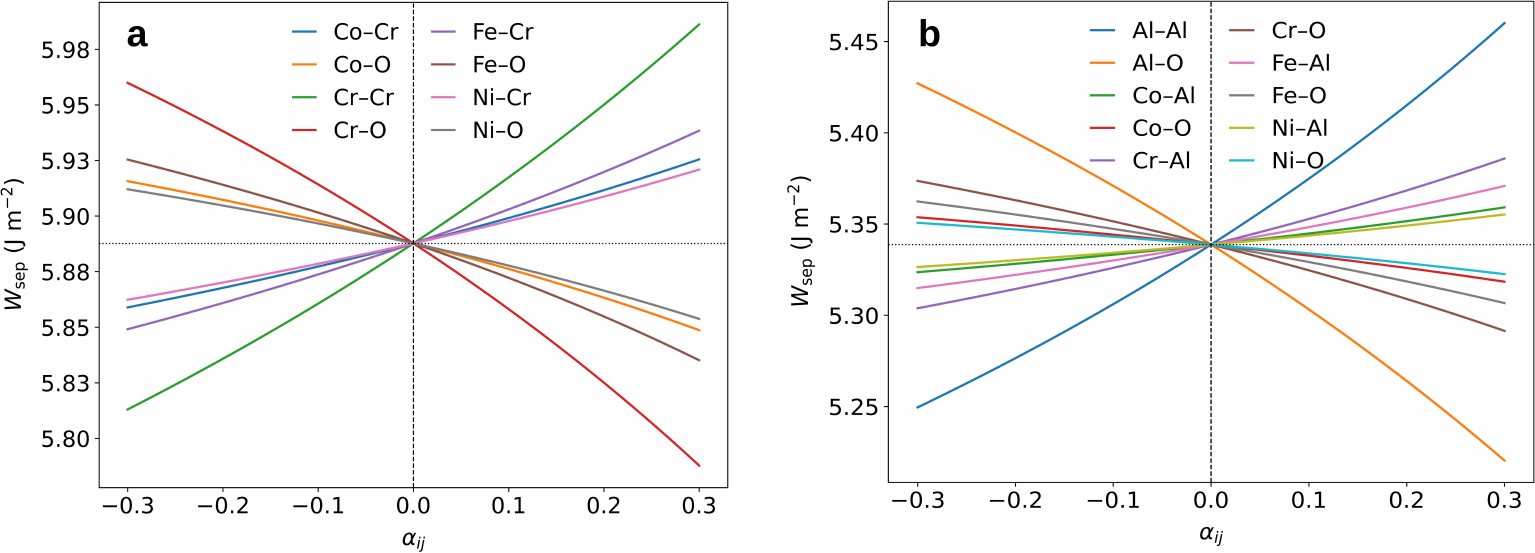}
    \caption{$W_{sep}$ as a function of $\alpha_{ij}$ of (a) CoCrFeNi-$\mathrm{Cr_2O_3}$, and (b) AlCoCrFeNi-$\mathrm{Al_2O_3}$ interfaces. An increase in Cr-O or Al-O pairs enhances adhesion, while Cr-Cr or Al-Al pairs drastically decrease adhesion. The order of influencing effect is Al-O, Cr-O, Fe-O, Co-O, and Ni-O, which is determined by $\Delta H_{ij}$.}
    \label{fig:pair_wise}
\end{figure}

To understand the effect of chemical ordering on interface adhesion, Figure \ref{fig:pair_wise} illustrates the variation of $W_{sep}$, as a function of the pair-probability parameter $\alpha_{ij}$ for all alloy-oxide elemental pairs. $\alpha_{ij}$ represents deviations from random interfacial contact statistics, which can be compared to short-range ordering in concentrated alloys. Values of $\alpha_{ij}$ correspond to random mixing, while negative and positive values indicate preferential or avoided interfacial contacts, respectively. Each curve corresponds to varying a single $\alpha_{ij}$ parameter in the range $-0.3 \leq \alpha_{ij} \leq 0.3$, while all other pairs remain in the random-mixing limit. It is apparent that the response of $W_{sep}$ to $\alpha_{ij}$ is continuous and approximately linear for all pairs considered. This shows that the pair-probability introduces a smooth and well-behaved perturbation to the interfacial energetics. No abrupt changes or non-physical instabilities are observed, confirming that the normalization of the pair probabilities preserves the mean-field form of the classical MAM.

An important thing to note is that the magnitude and sign of the response depend strongly on the chemical identity of the interacting elements. Pairs involving strong metal-oxygen interactions exhibit the largest slopes and therefore the strongest influence on $W_{sep}$. In particular, pairs such as Al-O and Cr-O show a clear positive contribution to $W_{sep}$ when $\alpha_{ij}$ becomes negative, indicating that an increased probability of forming these contacts strengthens interfacial adhesion. As captured in our previous work, this behavior is consistent with the well-known affinity of Al and Cr for oxygen and their role in stabilizing oxide scales through strong chemical bonding \cite{boakye2025effect,boakye2025first,zhang2025microstructure}.

Pairs involving Ni and Co display a noticeably weaker response. The Ni-O or Co-O curves are relatively flat compared to those of Cr-O and Al-O, indicating that modifying the probability of these contacts has only a limited effect on the work of separation. This explains the comparatively weaker chemical interaction between Ni, Co, and O, which contributes less to the overall interfacial bonding. As a result, changes in Ni- or Co-related contact statistics are largely absorbed by normalization of the pair probabilities and do not significantly change the net adhesion. 

Furthermore, pairs associated with metallic interactions across the interface, or with elements that do not form strong bonds with oxygen, show similarly weak or negligible effects. These curves are closely clustered and exhibit minimal slopes, demonstrating that variations in such contact probabilities do not play a dominant role in the determination of $W_{sep}$. This separation of strongly and weakly influential pairs highlights that adhesion in HEA-oxide systems is mainly controlled by a subset of chemically active metal-oxygen interactions rather than by all possible interfacial contacts. Another significant observation is that Al-Al and Cr-Cr pairs formed between Al (Cr) atoms in the matrix and the oxide negatively affect $W_{sep}$ compared to Cr-O and Al-O pairs. This could be explained by considering the dilute-limit enthalpies of Cr-O (-578 kJ/mol) and Al-O (-838 kJ/mol), compared to 0 kJ/mol for Cr-Cr or Al-Al. Hence, the formation of these pairs significantly reduces the total interfacial energy and consequently positively affects $W_{sep}$.

\subsection{Effect of temperature on segregation and adhesion}
While the present formulation evaluates segregation and interfacial adhesion primarily on the basis of enthalpic contributions, oxidation and scale formation occur at elevated temperatures, where entropic effects may also contribute to interfacial thermodynamics. In general, the temperature dependence of segregation is governed by the change in Gibbs free energy
\begin{equation}
    \Delta G = \Delta H - T\Delta S
\end{equation}
where for HEAs, configurational and vibrational entropy terms can partially offset the enthalpic driving force. However, for the REs considered here, the dominant contributions arise from strong chemical interactions between solutes and oxygen, particularly for Hf-O and Y-O bonding, whose dilute interaction enthalpies are highly exothermic. These interaction energies are typically on the order of several hundred to more than one thousand kJ/mol, substantially exceeding the expected magnitude of configurational entropy contributions at typical oxidation temperatures. Consequently, while temperature may modify the magnitude of segregation and adhesion, it is not expected to change the qualitative hierarchy of segregation tendencies or the relative effectiveness of REs over S. Instead, an increase in temperature primarily enhances diffusion kinetics, enabling the thermodynamically favored segregation predicted here to be realized more readily. Therefore, the present enthalpy-dominated treatment provides a physically meaningful approximation for high-temperature behavior, capturing the principal chemical driving forces that govern segregation-controlled adhesion in oxide-forming HEAs.

\subsection{Recommendation and future works}
The MAM serves not only as a predictive tool for oxide–metal adhesion but also as a generic thermodynamic framework for complex interfaces. By reformulating surface fractions and introducing an interfacial pair-probability formalism that captures deviations from random contact statistics, the model explicitly links local chemical ordering to macroscopic interfacial energetics. Because grain boundaries (GB) can be interpreted within MAM as internal interfaces characterized by broken coordination and altered contact statistics, the same idea can be applied directly to GB segregation. In such a treatment, $\alpha_{ij}$ would describe preferential or avoided contacts across boundary planes, while dilute-limit interaction enthalpies would quantify the competition between solute–matrix and solute–solute bonding. This allows a thermodynamic description of single and co-segregation at GB, including the trapping of deleterious species such as S by REs, without resorting to exhaustive atomistic sampling. Moreover, the smooth and nearly linear dependence of adhesion on contact statistics demonstrated here suggests that GB cohesion and embrittlement can be continuously mapped as functions of composition and local chemical order. Hence, the framework offers a scalable approach to predictive modeling of segregation-controlled cohesion in HEAs. This bridges the gap between atomistic first-principles calculations and mesoscale mechanical behavior, opening opportunities for rapid discovery of alloy compositions in which GB chemistry dictates long-term performance.

\section{Conclusion}
In this study, the macroscopic atom model (MAM) was extended to provide a quantitatively consistent and transferable description of segregation-mediated oxide-metal adhesion in high-entropy alloys. By introducing compositionally corrected surface fractions and an interfacial pair probability formalism, the model enables continuous prediction of segregation energies, surface energies, and the work of separation across alloy compositions, solute concentrations, and co-segregation. Moreover, the model reproduces the same segregation hierarchy as that of DFT, with Y followed by Hf and then S, and consistently predicts stronger segregation to alumina interfaces than to chromia interfaces for all reactive elements, in agreement with atomistic and experimental trends. For surface thermodynamics, MAM captures the nonlinear response to solute addition, predicting a reduction in surface energy with Y and S and an increase with Hf, with magnitudes comparable to DFT trends. For adhesion, the model produces work of separation values that mirror the behavior of DFT, increasing from about 4.9 $\mathrm{Jm^{-2}}$ for the clean CoCrFeNi $\mathrm{Cr_2O_3}$ interface to roughly 6.0 $\mathrm{Jm^{-2}}$ with Hf, while sulfur reduces $W_{sep}$ to around 4.05 $\mathrm{Jm^{-2}}$, and co segregation of reactive elements with sulfur partially restores adhesion. The prediction of the effect of oxide dispersoids was found to be positively effective for $HfO_2$, 6.11 $\mathrm{Jm^{-2}}$ in NiCoCrAl HEA, while a detrimental effect of S was observed, 3.26 $\mathrm{Jm^{-2}}$ for the NiAl superalloy.  Beyond reproducing these benchmarks, the model reveals that adhesion is primarily controlled by a small subset of chemically active metal-oxygen contacts, particularly Cr-O and Fe-O, and that systematic variation in their pair probabilities produces a smooth, nearly linear response in $W_{sep}$. These results demonstrate that the model is not only qualitatively correct but also quantitatively predictive, and provides a computationally efficient and physically interpretable path for screening segregation-controlled adhesion in chemically complex alloys beyond the limits of exhaustive first-principles calculations.

\section*{Acknowledgments}
This research was supported by the NSERC Alliance International Catalyst (ALLRP 592696-24), Canada, and the use of a high-performance computing system at the University of Manitoba and the Research Alliances of Canada. 

\section*{CRediT authorship contribution statement}
\textbf{Dennis Boakye}: Writing – review and editing, Writing – original draft, Visualization, Validation, Methodology, Investigation, Formal analysis, Data curation. \textbf{Chuang Deng}: Writing – review and editing, supervision, software, resources, project administration, investigation, Fund Acquisition, conceptualization.

\section*{Declarations}
The authors declare that they have no known competing financial interests or personal relationships that could have influenced the work reported in this paper.

\section*{Data availability}
The Python code is available on reasonable request.

\section*{Supplementary information}
The supplementary information referenced in the main text is attached as the Supplemental material.

\bibliographystyle{elsarticle-num} 
\biboptions{sort&compress}
\bibliography{references}
\end{document}